\documentclass[twocolumn]{webofc}
\usepackage[varg]{txfonts}   
\wocname{EPJ Web of Conferences}
\woctitle{International Symposium on Very High Energy Cosmic Ray Interactions 2014}
\begin{document}
\title{Atmospheric Lepton Fluxes}
%
%

\author{Thomas K. Gaisser\inst{1}\fnsep\thanks{\email{gaisser@bartol.udel.edu
}} 
}

\institute{Bartol Research Institute, Department of Physics and Astronomy\\
University of Delaware, Newark, DE 19716 (USA)
          }

\abstract{This review of atmospheric muons and neutrinos emphasizes the high
energy range relevant for backgrounds to high-energy neutrinos of astrophysical
origin.  After a brief historical introduction, the main distinguishing features
of atmospheric $\nu_\mu$ and $\nu_e$ are discussed, along with the implications
of the muon charge ratio for the $\nu_\mu/\bar{\nu}_\mu$ ratio.  Methods to
account for effects of the knee in the primary cosmic-ray spectrum and the energy-dependence
of hadronic interactions on the neutrino fluxes are discussed and illustrated
in the context of recent results from IceCube.
A simple numerical/analytic method is proposed for systematic investigation
of uncertainties in neutrino fluxes arising from uncertainties in the primary 
cosmic-ray spectrum/composition and hadronic interactions. 
}
\maketitle
\section{Introduction}
\label{sec1}
Atmospheric leptons are of current interest in two contexts, as a beam
for the study of neutrino oscillations and the mass hierarchy
(energy range $100$~MeV$\rightarrow$TeV)
and as the background in the search for high energy neutrinos of
astrophysical origin (energy range $100$~GeV$\rightarrow$PeV).  The emphasis
of this paper is on the higher energy range, motivated by the discovery
by IceCube~\cite{Aartsen:2013jdh,Aartsen:2014gkd} of neutrinos of 
extraterrestrial origin at very high energy
above the background of atmospheric neutrinos.  In view of this discovery
it is important to understand the atmospheric neutrino spectrum as
precisely as possible, not only to understand the backgrounds,
but also in order to learn how the spectrum
of the astrophysical component extends to lower energy.

\section{History}
\label{sec2}
It is well known that the idea of using Cherenkov light in a large volume of water to
detect interactions of high-energy neutrinos was suggested in 1960 by
Markov~\cite{Markov}, by Greisen~\cite{Greisen:1960wc} 
and by Reines~\cite{Reines:1960we}.
Markov was interested in using atmospheric neutrinos to 
measure the energy dependence of the neutrino cross section.
He speculates about the existence of a vector boson intermediary
and the question of whether there are two different kinds of neutrinos
related to the electron and the muon.
At the end of his review, ``Cosmic Ray Showers,''
Greisen speculated briefly about detecting extraterrestrial neutrinos.
Reines' review of ``Neutrino Interactions'' in the same volume 
devotes a section to ``Cosmic and Cosmic Ray Neutrinos.''
By ``cosmic neutrinos'' he means neutrinos produced by cosmic rays
in or near their distant sources, while ``cosmic-ray neutrinos'' refers to
atmospheric neutrinos produced by interactions of cosmic rays at Earth.
After noting the potential of cosmic neutrinos as a probe of the origin of cosmic rays
and noting the ignorance of what flux to expect,
he writes, ``The situation is somewhat simpler in the case of cosmic-ray neutrinos: 
they are both more predictable and of less intrinsic interest.''
He goes on to estimate a detection rate of $\sim$1 atmospheric neutrino
per day in 5000 m$^3$ of water.

Perhaps less well known is the early (1961) paper~\cite{ZatKuz} by Zatsepin
and Kuz'min in which the essential phenomenology of atmospheric neutrinos
is derived, including the important role of kaons relative to pions
as parents of neutrinos.  Their papers include the production of
neutrinos from decay of muons, and they describe the strong angular dependence
at high energy, which is a consequence of the $\sec\theta$ dependence
of the ratio of decay to interaction above the critical energies of
the mesons.  They also explain that charged kaons are more efficient
producers of neutrinos than charged pions because of the higher mass of
the kaon relative to the muon and the shorter lifetime of the kaon.
In the 60's and 70's Volkova and Zatsepin published a series of papers
refining the calculations, as described in the paper of Volkova~\cite{Volkova:1980sw},
which remains a standard reference for fluxes of atmospheric neutrinos.

\section{Overview of neutrinos and muons}
\label{sec3}
The linear development of the hadronic cascade 
in the atmosphere is described
by a system of equations of the form
\begin{eqnarray}
\label{MasterEqn}
{{\rm d}N_i(E_i,X)\over {\rm d}X} &=&-{N_i(E_i,X)\over\lambda_i} 
                                             - {N_i(E_i,X)\over d_i} \\ 
& & +\Sigma_{j=i}^J\int_E^\infty\,{F_{ji}(E_i,E_j)\over E_i}\,
{N_j(E_j,X)\over\lambda_j}\,{\rm d}E_j, \nonumber
\end{eqnarray}
where $N_i(E_i,X)$d$E_i$ is the flux of particles of type $i$ at
slant depth $X$ in the atmosphere with energies in the 
interval $E$ to $E+{\rm d}E$~\cite{Gaisser:1990vg,Lipari:1993hd}.
The probability for a particle of type $j$ to interact in d$X$ is
d$X/\lambda_j(E_j)$, where $\lambda_j$ is the corresponding interaction length (in g/cm$^2$).
Similarly,
d$X/{d_j(E_j)}$ is the probability that a particle of type $j$ decays in d$X$.
The function $F_{ji}(E_i,E_j)$ is
\begin{equation}
F_{ji}(E_i,E_j)\equiv E_i{{\rm d}n_i(E_i,E_j)\over {\rm d}E_i}, 
\label{inclusive}
\end{equation}
where d$n_i$ is the number of particles of type i produced on average
in the energy bin d$E_i$ around $E_i$ per collision of an incident particle 
of type $j$.  Depending on the boundary condition, the solution of Eq.~\ref{MasterEqn}
can describe either a single air shower or the inclusive flux
of a particular particle type in the atmosphere.  
In the latter (inclusive) case, the boundary
condition is $N_N(X=0)=\phi_N(E_N)$ and $N_i(X=0)=0$ for $i\ne N$.
Here $\phi_N(E_N)$ is the spectrum of nucleons at the top of the atmosphere.
Correspondingly, the solutions $N_i(E_i,X,\theta)$ give the inclusive rates of particles
of type $i$ per unit area at depth $X$ independently of whether or not there are
particles nearby.  In contrast, for the air shower boundary
condition ($N_A(E_A,X=0)=\delta(E=E_0)$) the solution in principle contains all
the correlations among the particles in the shower.

\begin{figure}
\centering
\includegraphics[width=8cm,clip]{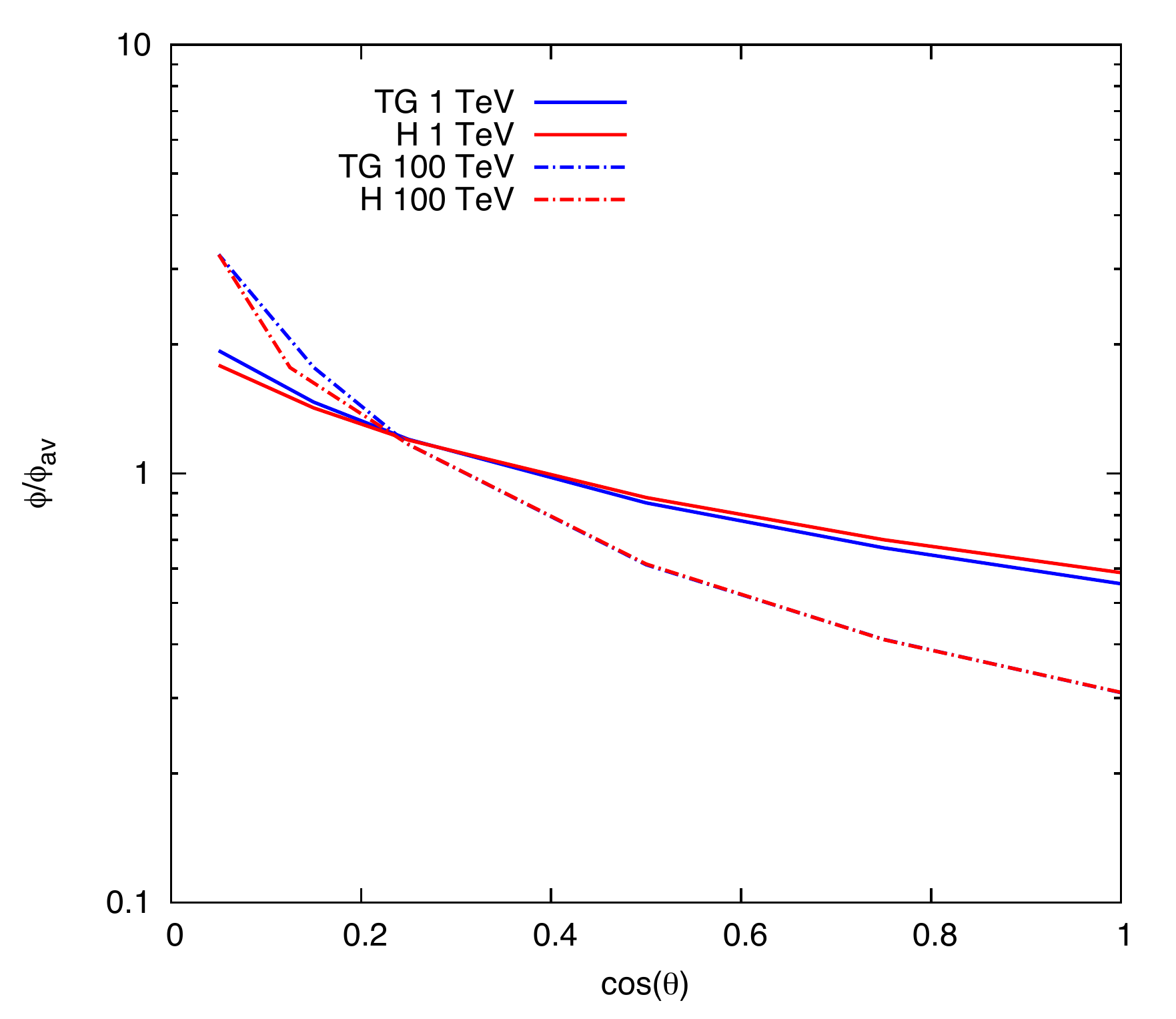}
\caption{Normalized angular distributions of muon neutrinos
at 1 TeV (solid lines) and 100 TeV (dashed lines).}
\label{fig:angular}       
\end{figure}

\subsection{Muon neutrinos}

For the inclusive boundary condition with 
a power-law spectrum of primary nucleons
and assuming scaling of the particle interactions
in the forward kinematic region, a good approximation to 
the solution of Eq.~\ref{MasterEqn} for the flux of $\nu_\mu+\bar{\nu}_\mu$ is
\begin{equation}
\phi_\nu(E_\nu) =  \phi_N(E_\nu) \times \sum_{i=1}^3
 \left({A_{i\nu}\over 1 + 
B_{i\nu}\cos^*\theta\, E_\nu / \epsilon_i}\right),
\label{angular}
\end{equation}
where the sum is over the contributions of charged pions,
charged kaons and charmed hadrons.  At low energy ($<100\,{\rm GeV}/\cos^*\theta$)
the contribution from decay of muons also needs to be added.
The asterisk on cosine of the zenith angle $\theta$ indicates the
correction necessary to account for the curvature of the Earth
for $\theta>60^\circ$~\cite{Volkova:1980sw}.  The quantity
$\epsilon_i/\cos^*\theta$ is the energy for meson $i$
above which re-interaction in the atmosphere becomes more likely
than decay.  At energies below the critical energy ($\epsilon_i$) for each
channel, most mesons decay, so the angular distribution of the corresponding
neutrinos is isotropic for an isotropic primary cosmic-ray spectrum.  In addition,
the neutrino spectrum is similar to the cosmic-ray spectrum.
For $E> \epsilon_i$ the energy spectrum steepens, first near
the vertical and at higher energy for more horizontal directions.
As a consequence, at high energy the intensity of neutrinos from
near the horizon is an order of magnitude larger that near the vertical,
as illustrated in Fig.~\ref{fig:angular}.
The short lifetime of charmed hadrons corresponds to $\epsilon_{\rm charm}\sim 10^7$~GeV,
so the small fraction of prompt neutrinos will be isotropic below
this energy.

The numerator of Eq.~\ref{angular} contains the spectrum-weighted moments
for production of secondary particles in particle collisions in
the atmosphere.  For example, for nucleon$\rightarrow K^+\rightarrow \nu_\mu$
\begin{equation}
A_{K^+\nu}=\frac{Z_{NK^+}}{1-Z_{NN}}\times Z_{K^+\nu},
\label{Afactor}
\end{equation}
where 
\begin{equation}
Z_{NK^+}(E)\,=\,\int_E^\infty\,{\rm d}E'\frac{\phi_N(E')}{\phi_N(E)}
\frac{\lambda_N(E)}{\lambda_N(E')}\frac{{\rm d}n_{K^+}(E',E)}{{\rm d}E}
\label{TIG-ZK}
\end{equation}
and $Z_{NN}$ has a similar definition in terms of the cross section
for a nucleon of energy $E^\prime$ to produce a secondary nucleon of 
energy $E$.  $Z_{K^+\nu}$ is the spectrum weighted moment of the decay
distribution.

Eq.~\ref{TIG-ZK} is a general definition given in Ref.~\cite{Gondolo:1995fq} 
that takes account of the
energy dependence of interaction cross sections and particle production
and allows for a general form of the primary spectrum.  For a power-law
primary spectrum with differential index $1+\gamma$, constant interaction
cross sections and particle production that depends only on the
ratio $x=E/E^\prime$, Eq.~\ref{TIG-ZK} simplifies to the more familiar,
energy-independent form,
\begin{equation}
Z_{NK^+}\,=\,\int_0^1\,x^\gamma \frac{{\rm d}n_{K^+}(x)}{{\rm d}x}.
\label{Zscaling}
\end{equation}

\subsection{Electron neutrinos}

Electron neutrinos come primarily from three-body decays
of kaons at energies high enough so that the contribution
from muons is small.  The intensity of atmospheric electron
neutrinos is approximately 5\% of muon neutrinos at high energy.  The contribution
of charm is the same for both $\nu_e$ and $\nu_\mu$, which means
that the fraction of prompt $\nu_e$ is significantly larger than
the fraction of prompt $\nu_\mu$.  The lower normalization of
$\nu_e$ from kaons also means that the contribution from
muon decay is relatively more important to higher energy
and lower zenith angle than for $\nu_\mu$.  A detailed calculation
of electron neutrinos at high energy, including the contribution from the
rare three-body decay of $K_S$, is given in Ref.~\cite{Gaisser:2014pda}.
Because of its short lifetime, the critical energy for $K_S$
is $120$~TeV.  As a consequence, the spectrum of neutrinos from
this channel is harder by one power of energy than that from
$K^\pm$ and $K_L$.  Above $100$ TeV the contribution from
the three channels is nearly equal because the lifetime is
inversely proportional to the total decay width and the branching
ratio is the ratio of the semi-leptonic width to the total width.

\begin{figure}
\centering
\includegraphics[width=8cm,clip]{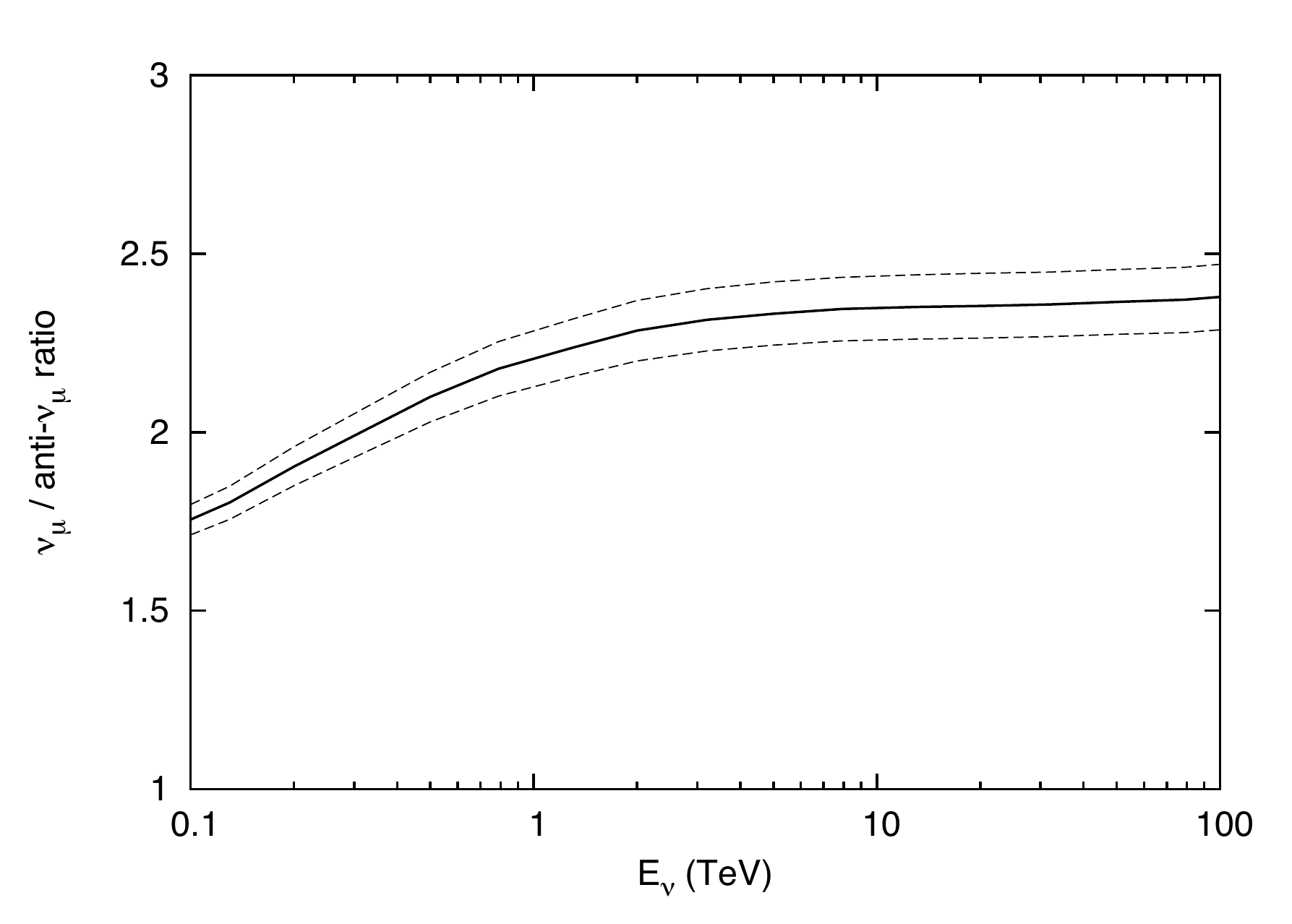}
\caption{Ratio of $\nu_\mu/\bar{\nu}_\mu$ inferred from the
 the $\mu^+/\mu^-$ ratio of OPERA~\cite{Agafonova:2014zia}.}
\label{fig:ratio}       
\end{figure}

\subsection{Muons and the $\mu^+/\mu^-$ ratio}

The flux of muons in the atmosphere is similar in form and closely related to
the flux of muon neutrinos, with three terms as in Eq.~\ref{angular}.  
The abundant and well-measured spectrum of
atmospheric muons is therefore an important benchmark for any calculation of neutrinos.
In fact, measurements of atmospheric muons~\cite{Sanuki:2006yd} are used to tune the
model for hadron production for one of the standard 
neutrino flux calculations~\cite{Honda:2006qj}.
The main difference between $\mu$ and $\nu_\mu$
is the kinematics of the two-body decays of charged pions and kaons.
Because the muon carries most of the energy in pion decay, and because
more pions are produced than kaons, most muons come from the pions, even
after accounting for the steep primary spectrum and the higher value
of $\epsilon_K$.

The signature of the kaon channel nevertheless shows up in an important way
in the charge ratio of muons, which increases from $\approx 1.28$
around $100$~GeV~\cite{Achard:2004ws} to 
$\approx 1.4$ at a TeV and above~\cite{Adamson:2007ww,Agafonova:2014zia}.
Although most muons come from $\pi^\pm \rightarrow\mu+\nu_\mu$,
the fraction from kaons increases above the energy at which
charged pions begin to prefer interaction to decay in the atmosphere.
The critical energy for pions is $\approx 115$~GeV compared to 
$\approx 850$~GeV for kaons.  There is an increase in
the muon charge ratio in this energy range because of the importance of associated production,
$p\rightarrow K^+\Lambda$, which makes the $\pm$ charge ratio higher for 
kaons than for pions.  The fluxes of $\mu^+$ and $\mu^-$ can be calculated
in a straightforward way by keeping track of positive and negative
meson fluxes separately~\cite{Gaisser:2012zz}.  The ratio depends both on the excess of
protons in the primary spectrum and on the spectrum-weighted moments
for the separate meson charge channels.
Using the calculation of Ref.~\cite{Gaisser:2012zz} to fit
the proton excess in the primary spectrum and the Z-factor, the OPERA group found
$\delta_0=(p_0-n_0)/(p_0+n_0)=0.61\pm 0.02$ and $Z_{pK^+} = 0.0086\pm 0.0004$
at a mean muon energy of $2$~TeV~\cite{Agafonova:2014zia}.  This result has implications for
the ratio $\nu_\mu/\bar{\nu}_\mu$ in the same energy region.  With the
value of $Z_{pK^+}$ from OPERA, the expected ratio for muon neutrinos increases from
$\nu_\mu/\bar{\nu}_\mu\approx 1.5$ at low energy to $\approx 2.3$ above a TeV, as
shown in Fig.~\ref{fig:ratio}.  

\section{Analytic and numerical calculations of atmospheric neutrinos}
\label{sec4}

A formal expression for the flux of neutrinos is
\begin{equation}
\phi_\nu(E_\nu,\theta) = \sum_A\int_{E_\nu}^\infty\,\phi_A(E_A)Y_\nu(A,E_\nu,E_A,\theta){\rm d}E_A,
\label{formal}
\end{equation}
where $\phi_A$ is the primary flux of nuclei with total energy $E_A$ and
$Y_\nu(A,E_\nu,E_A,\theta)$ is the yield of neutrinos from a given
primary nucleus.  A Monte Carlo evaluation of this
integral is the standard approach for detailed calculations of
the fluxes of atmospheric muons and neutrinos~\cite{Honda:2006qj,Barr:2004br}.
The results can be re-weighted to correspond to any desired description
of the primary spectrum and composition, including 
the effects of geomagnetic cutoffs at different locations.  
It is straightforward to include muon energy loss and decay.
The flux can be extended to
high energy with good statistics by forcing all mesons to decay and recording
the fractional probability that the decay would actually have occurred.
In the calculation
of Ref.~\cite{Barr:2004br}, the yields were calculated by Monte Carlo for five representative
nuclei (p, He, N, Si, Fe) at ten primary energies per decade, equally spaced in $\log(E_A)$.
The neutrino flux bins were filled with the appropriate fractional weights in logarithmically
spaced bins of neutrino energies.  For both Refs.~\cite{Honda:2006qj,Barr:2004br}
the calculations extend only to $E_\nu=10$~TeV.

Analytic and numerical calculations are a useful alternative to the
Monte Carlo approach because of the insight
they provide into which aspects of the physics are most important for
different features of the spectra of atmospheric leptons.  
A simple, fast analytic/numerical routine is also suitable for
systematic exploration of uncertainties in the lepton fluxes
due to lack of knowledge of 
the primary spectrum/composition and to the hadronic interactions.  Two approaches were
described at this conference, the numerical integration of the
coupled matrix of equations~\cite{Anatoli} and the iterative scheme~\cite{Naumov}
of Sinegovskaya et al.~\cite{Sinegovskaya:2014pia}.  The approach developed
in this paper uses energy-dependent Z-factors to account for energy dependence
of hadronic interactions and non-power-law behavior of the primary spectrum.

\begin{figure*}
\centering
\includegraphics[width=8.5cm,clip]{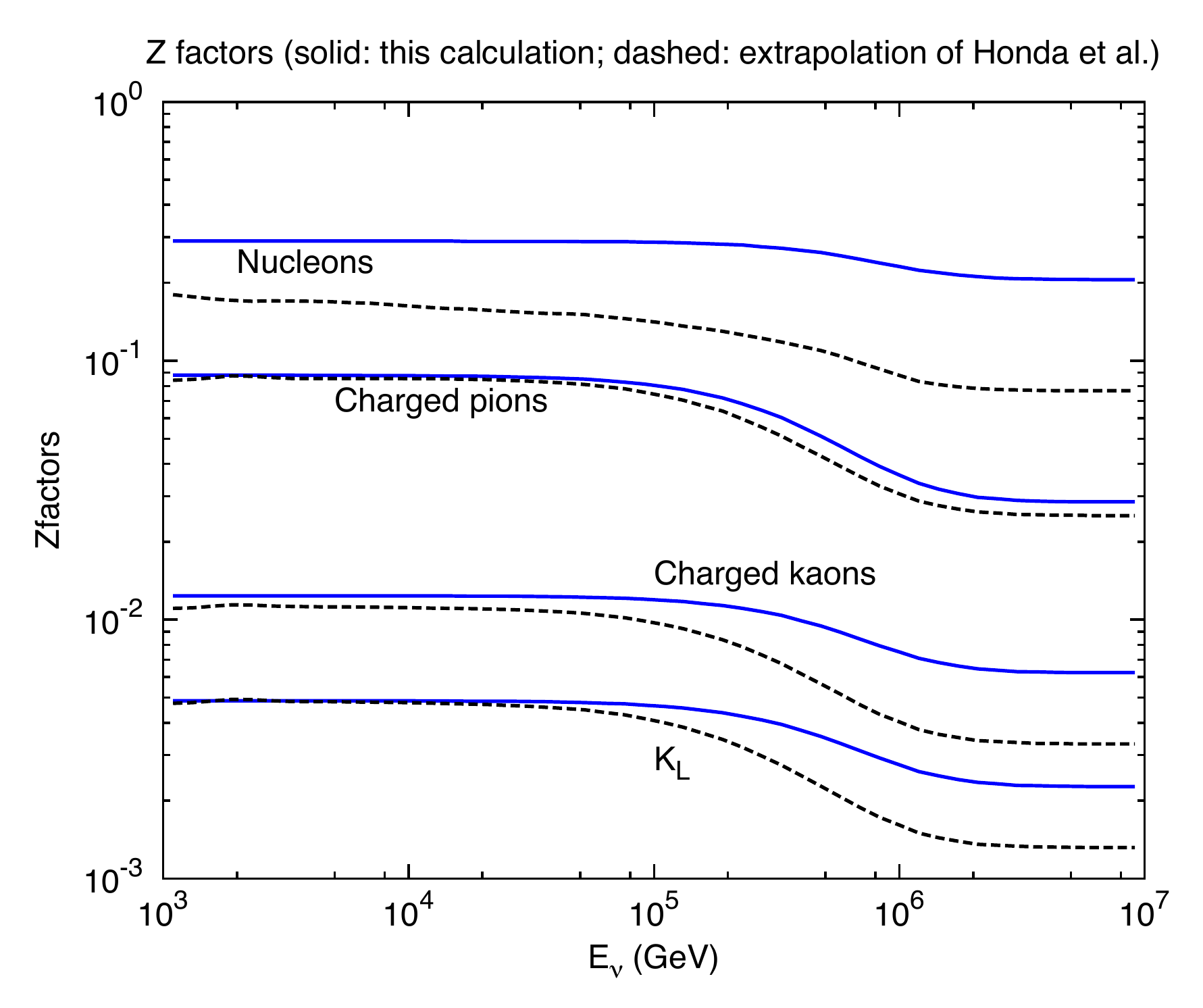}\,\includegraphics[width=7.5cm,clip]{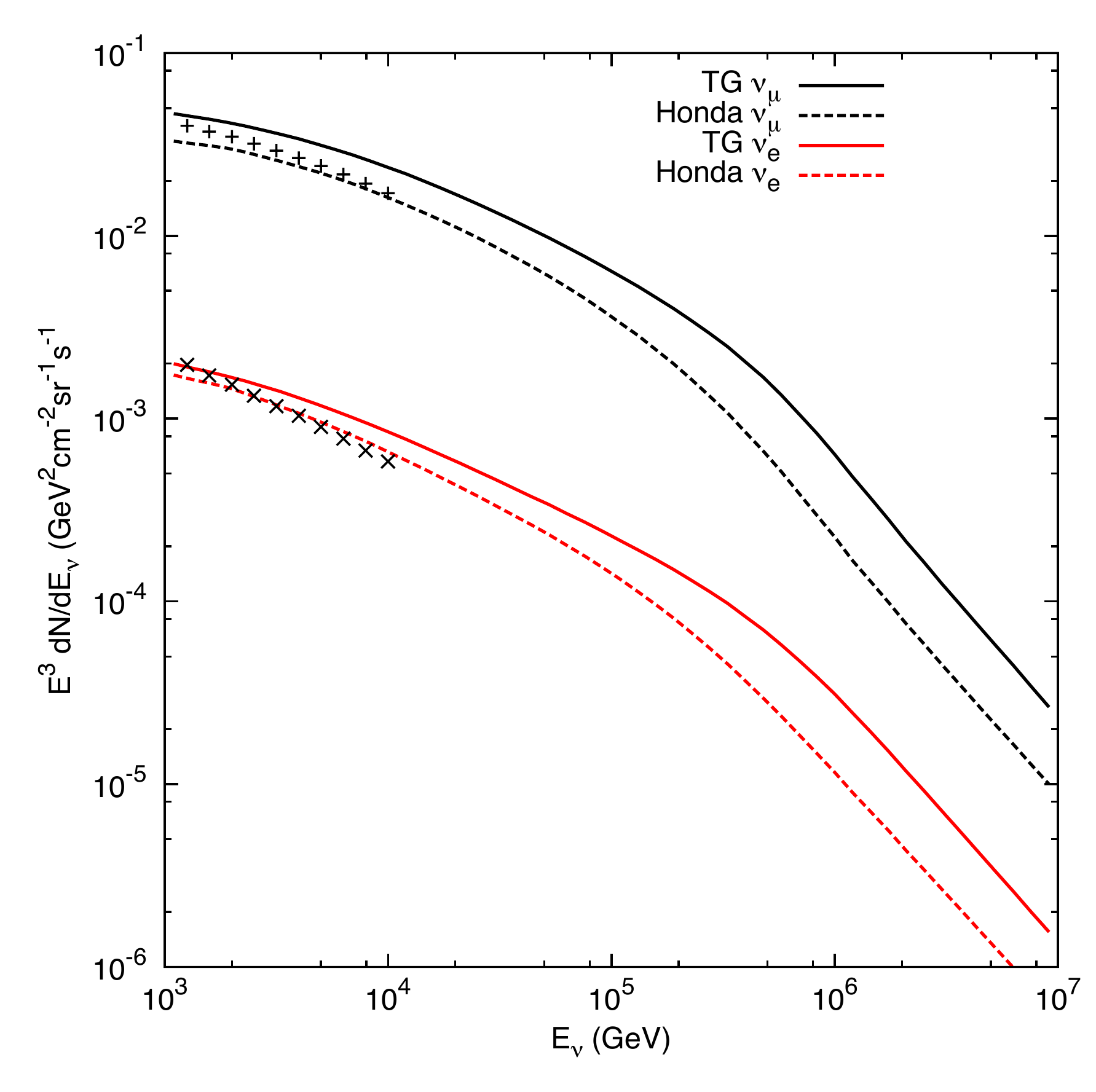}
\caption{Left: Energy-dependent Z-factors from Ref.~\cite{Gaisser:2001sd} (solid lines)
compared to those of Ref.~\cite{Honda:2006qj} (broken lines).  Right:
Spectrum of $\nu_\mu+\bar{\nu}_\mu$ and $\nu_e+\bar{\nu}_e$ for two
models of particle production calculated with the primary spectrum 
of Ref.~\cite{Gaisser:2012zz}, which includes the knee.
The crosses are the tabulated neutrino fluxes averaged over the sky from 
Ref.~\cite{Honda:2006qj} (see text for discussion).}
\label{fig:Zfactors}       
\end{figure*}

\subsection{Energy-dependent Z-factors}
In their paper on neutrinos and muons from charm decay, 
Gondolo, Ingelman \& Thunman~\cite{Gondolo:1995fq}
showed that the simple Eq.~\ref{angular} valid for a power-law primary spectrum
and scaling can be adapted to include energy dependence.  With the generalized definition
of the Z-factors as in the example of Eq.~\ref{TIG-ZK}, the algebraic solutions
(Eq.~\ref{angular}) correctly account both for the energy dependence of meson production and
for the non-power-law behavior of the primary spectrum, 
provided that the energy dependences are sufficiently
gradual.  This scheme was used in Ref.~\cite{Fedynitch:2012fs} and Ref.~\cite{Gaisser:2013ira} 
to account for the 
effect of the knee in the primary spectrum.  Here I also give an example of the
comparison of two different interaction models.
Evaluating the energy-dependent Z-factors as in Eq.~\ref{TIG-ZK} requires
a representation of the production distribution
for each particle considered.  Using the simple, two-parameter forms of Ref.~\cite{Gaisser:2001sd}
is sufficient, provided the parameters are tabulated at a sufficiently fine grid
of energies.  The production spectrum for $j\rightarrow i$ is
\begin{equation}
\frac{{\rm d}n_{i,j}(E_i,E_j)}{{\rm d}x}=c_{i,j}\frac{(1-x)^{p_{i,j}}}{x},
\label{an1}
\end{equation}
with $x=E_i/E_j$ and the two energy-dependent parameters $c_{i,j}(E_j)$ and $p_{i,j}(E_j)$.
One way to characterize a hadronic event generator is by a set of
spectrum weighted moments calculated at each beam energy for
a range of power-law spectra.  Then from
\begin{equation}
Z_{i,j}(E_j,\gamma)=\int_0^1\,x^\gamma\frac{{\rm d}n_{i,j}(E_j)}{{\rm d}x}{\rm d}x
\label{an2}
\end{equation}
the energy-dependent parameters can be evaluated at each energy as
\begin{equation}
p_{i,j}(E_j) = \frac{Z_{i,j}(E_j,1)}{Z_{i,j}(E_j,2)}-2
\label{an3}
\end{equation}
and
\begin{equation}
c_{i,j}(E_j) = (p_{i,j}(E_j)+1)\times Z_{i,j}(E_j,1).
\label{an4}
\end{equation}
Having determined the parameters for particle production,
The next step is to evaluate the spectrum-weighted moments
at each energy for an arbitrary spectrum using
\begin{equation}
Z_{i,j}(E) = \int_0^1\frac{{\rm d}x}{x}\frac{\phi_p(E/x)}{\phi_p(E)}\frac{{\rm d}n_{i,j}(E,E/x)}{{\rm d}x}.
\label{an5}
\end{equation}
(For simplicity, the factor $\lambda(E)/\lambda(E/x)$ in Eq.~\ref{TIG-ZK} has been approximated
as unity here because the interaction cross section varies slowly over
the  range of $x$ near $1$ that is important for the integral over the steep primary spectrum.)

For illustration, two hadronic interaction models are compared here.
One is a very simple energy-independent extrapolation of the parameters from Ref.~\cite{Gaisser:2001sd},
which correspond to the Z-factors of Ref.~\cite{Gaisser:1990vg} for $\gamma=1.7$.  In this case,
the only energy dependence in the $Z_{i,j}(E)$ is from the knee of the spectrum,
as in Ref.~\cite{Gaisser:2013ira}.  For comparison, the more realistic case $Z_{i,j}(E)$
corresponding to the hadronic interaction model of Ref.~\cite{Honda:2006qj} are used.
The energy-dependent Z-factors (Eq.~\ref{an5}) are shown in Fig.~\ref{fig:Zfactors}~(left)
evaluated with the H3A primary spectrum of Ref.~\cite{Gaisser:2012zz}.  The corresponding
fluxes of $\nu_\mu+\bar{\nu}_\mu$ and $\nu_e+\bar{\nu}_e$ are shown in Fig.~\ref{fig:Zfactors}~(right),
and the angular distributions at two energies are shown in Fig.~\ref{fig:angular}.
The analytic calculations here do not include neutrinos from decay of muons, and
the tabulated data from Ref.~\cite{Honda:2006qj} are above the calculation at low
energy.  In addition, the tabulated values were calculated with a slightly different
energy spectrum, as noted in the discussion of Fig.~\ref{fig:primary} below.

\begin{figure}
\centering
\includegraphics[width=8cm,clip]{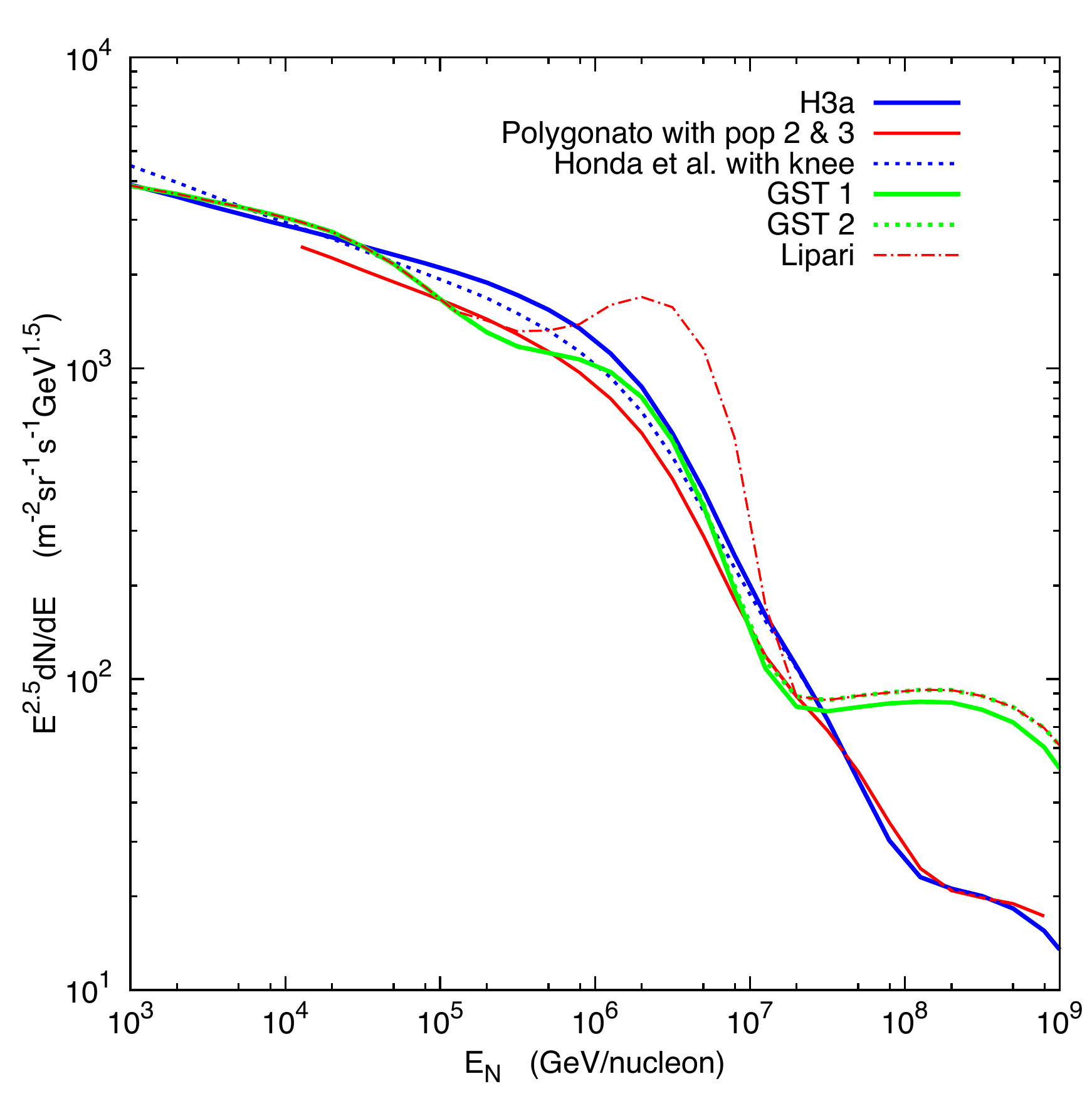}
\caption{Primary spectrum of nucleons for several
models of primary cosmic-ray composition and energy spectrum.
(See text for references and discussion of the different models.)
}
\label{fig:primary}       
\end{figure}
\subsection{Primary spectrum}

For an analytic or numerical approximation as in Eq.~\ref{angular}
the primary spectrum enters the calculation as the spectrum of nucleons per GeV/nucleon.
This superposition approximation neglects collective effects in nuclei, but correctly
captures the kinematics of meson production in nucleon-nucleon collisions.
We are interested in an energy range that spans direct measurements of the primary
spectrum below 100 TeV/nucleon and air shower measurements to the knee and beyond.
Fits to the spectrum attempt to extrapolate the direct measurements into the 
air-shower regime where the spectrum is generally presented as an all-particle
spectrum in energy per nucleus.  One common assumption in fitting the data
is to assume that the primary spectrum depends only on magnetic rigidity~\cite{Peters}.
The rationale is that both propagation and acceleration depend on motion in magnetic
fields.  Several such representations
of the nucleon spectrum are shown in Fig.~\ref{fig:primary}.

The Polygonato model~\cite{Hoerandel:2002yg} is an example in which each
element is assigned a power-law index based on direct measurements at low energy
and then suppressed above an energy $E_A=ZeR_{\rm cut}$, where $R_{\rm cut}$,
the cutoff rigidity, is tuned to fit the knee of the cosmic-ray spectrum.  
In Ref.~\cite{Gaisser:2012zz}, parameters for a model (H3a) 
with the standard five nuclear groups
(p, He, CNO, Si, Fe) and three populations are given.  Two galactic populations
have cutoffs of $R_1=4$~PV (for the knee) and $R_2=30$~PV.  The cutoff for
the extragalactic population depends on what composition is assumed.  Other
parameterizations (GST1 and GST2) based on a different fit to direct measurements and guided
by measurements of average primary composition from air showers measurements
are given in Ref.~\cite{Gaisser:2013bla}.  The Polygonato model
shown in Fig.~\ref{fig:primary} is a modified version with 5 nuclear components
at low energy and populations 2 and 3 of H3a.  The red dashed line 
 is a version of GST2 modified by Lipari~\cite{Lipari:2013taa} to include
an unrealistically high fraction of protons in order to estimate the maximum possible
atmospheric neutrino flux above 100 TeV.  The dotted line is the spectrum
used in the calculation of the neutrino flux by Honda et al.~\cite{Honda:2006qj}
modified to include the knee as in H3a.  This parameterization is the high helium
version of a parameterization and extrapolation of direct measurements from
Ref.~\cite{Gaisser:2001jw}.  The $\sim$30\% range in the knee region in Fig.~\ref{fig:primary}
translates into a comparable uncertainty for the atmospheric neutrino flux
above $\sim$30 TeV.

\begin{figure}
\centering
\includegraphics[width=8cm,clip]{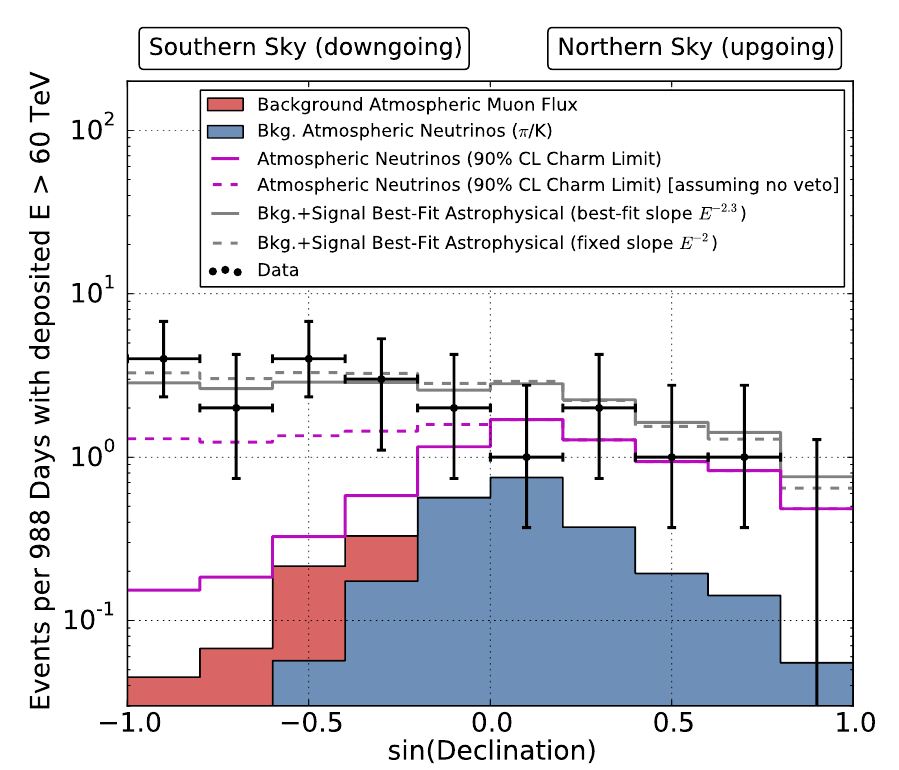}
\caption{Angular distribution of neutrino events with $E>60$~TeV in IceCube.
(From Ref.~\cite{Aartsen:2014gkd}, see text for discussion.)
}
\label{fig:HESEangular}       
\end{figure}

\section{Neutrino yields and neutrino self-veto}
\label{sec5}

The High Energy Starting Event (HESE) analysis in IceCube~\cite{Aartsen:2013jdh,Aartsen:2014gkd}
defines a veto region in the outer parts of the deep array.  Events are selected that start
in the fiducial volume inside the veto region.  The analysis also sets a very high
energy threshold by requiring an amount of light equivalent to approximately 30 TeV or more
of energy deposition in the detector.  A key feature of this analysis is that
high-energy atmospheric neutrinos from above are excluded from the event sample
if they are accompanied at the detector by a muon from the same shower that triggers the veto.
Evaluation of the veto probability is
an important case where the forced decay scheme for Monte Carlo evaluation
of Eq.~\ref{formal} at high energy does not work.   The reason is that
in this case it is essential to keep the correlation between the
neutrino and the high-energy muons in the same shower which provide the veto.

\begin{figure*}
\centering
\includegraphics[width=8cm,clip]{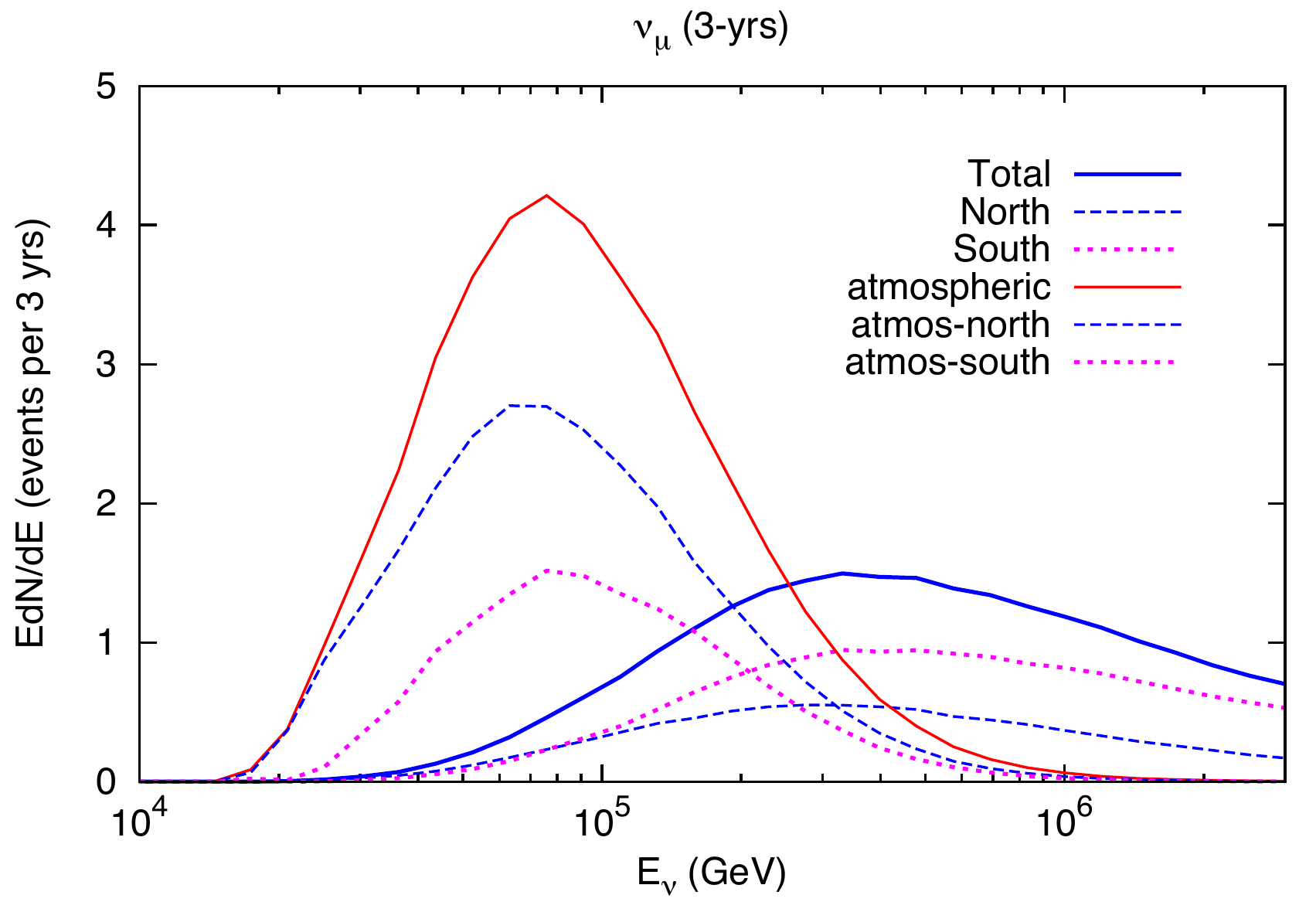}\,\includegraphics[width=8cm,clip]{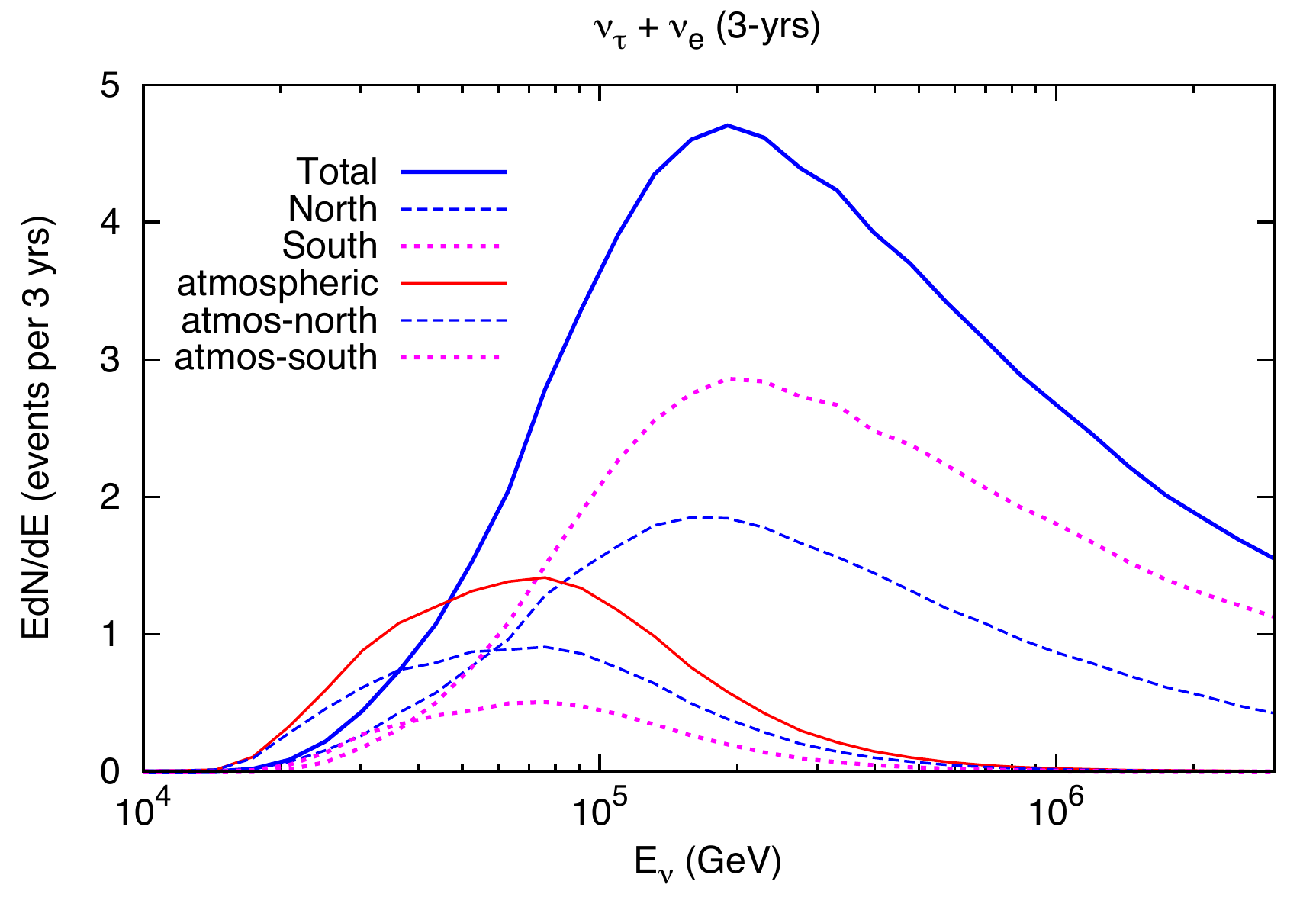}
\caption{Response functions showing the distributions of true neutrino energies
that give rise to events in the IceCube HESE analysis  (left, $\nu_\mu+\bar{\nu}_\mu$; 
right, $\nu_e+\bar{\nu}_e+\nu_\tau+\bar{\nu}_\tau$).}
\label{fig:nu-response}       
\end{figure*}

The probability that a muon produced in the same decay as a $\nu_\mu$ reaches
the deep detector with a minimum energy can be calculated analytically~\cite{Schonert:2008is}.
This calculation applies only to muon neutrinos.  In general, an atmospheric neutrino 
can also be accompanied by a muon from a different branch of the same air shower in which
the neutrino was produced.  Such a generalized self-veto can be applied to electron
neutrinos and to prompt neutrinos of either flavor.  A procedure for evaluating
the generalized self-veto~\cite{Gaisser:2014bja} consists of 
evaluating an expression similar to Eq.~\ref{formal} but with an extra factor in the
integrand:
\begin{eqnarray}
\nonumber
\phi_\nu(E_\nu,\theta\,|\,0) &=& \sum_A\int_{E_\nu}^\infty{\rm d}E_A\phi_A(E_A)Y_\nu(A,E_\nu,E_A,\theta) \\ 
& \times & \exp\{-N_\mu(E_A,E_\mu>E_{\rm min}(\theta))\}.
\label{generalized}
\end{eqnarray}
The exponential factor is the Poisson probability that a shower generated by a primary nucleus with
energy $E_A$ has no muons with $E_\mu>E_{\rm min}(\theta)$, the minimum muon energy at production
needed for the muon to reach the detector with sufficient residual energy to be detected.
A new set of parameterizations analogous to the Elbert formula~\cite{Elbert} was obtained from
simulations with Sibyll 2.1~\cite{Ahn:2009wx} to use for the yields in Eq.~\ref{generalized}.
The yields for both neutrinos and muons have the integral form
\begin{equation}
N_i(>E_i,A,E_A,\theta)=C_i\frac{A}{E_i\cos^*\theta}x^{-p_1}(1-x^{p_3})^{p_2}.
\label{params}
\end{equation}
In the original Elbert formula for muons, $p_3=1$.  In the expression for prompt neutrinos
the factor $1/E_i\cos^*\theta$ is omitted.  The differential yields are 
$$ Y_i(E_i)=\frac{{\rm d}N_i(>E_i)}{{\rm d}E_i}. $$  Then
the passing rate for atmospheric neutrinos of energy $E_\nu$ entering
the detector with zenith angle $\theta$ and interacting inside is
\begin{equation}
\frac{\phi(E_\nu,\theta\,|\,0)}{\phi(E_\nu)}
\label{passing}
\end{equation}

An IceCube analysis that makes extensive use of the atmospheric neutrino self-veto
was presented at this conference by J. van Santen, 
since published as Ref.~\cite{Aartsen:2014muf}.
The importance of the veto for the discovery of neutrinos in IceCube is illustrated by
Fig.~\ref{fig:HESEangular}, reproduced from the supplemental material of the IceCube
publication~\cite{Aartsen:2014gkd} and shown at this conference in a presentation by
G. Binder.  The pink histograms in the figure show the angular distribution
of prompt neutrinos from decay of charmed hadrons, which would likely be the dominant
contribution to the atmospheric background for $E_\nu>60$~TeV.  
The dashed line shows the shape before the veto, while the solid line 
is the expectation after the veto.   The angular axis is labeled in declination.
At the South Pole the zenith angle $\theta$ is related to declination $\delta$
by $\cos\theta=-\sin\delta$.  The veto suppresses the downward atmospheric
background by 50\% or more for zenith angles $\theta<70^\circ$, a shape
completely different from the data.

\section{Rates in IceCube}
\label{sec6}

Effective areas in the IceCube HESE analysis are 
given for three flavors of neutrinos separately for the Northern and Southern hemispheres
in the supplementary material of Ref.~\cite{Aartsen:2013jdh}.  Expected rates for
flavor $i$ are
obtained by the convolution of the neutrino spectrum with the corresponding effective area,
\begin{equation}
{\rm Rate}_i\,=\,\iint\,A_{i,{\rm eff}}(E_\nu)\phi_i(E_\nu){\rm d}E_\nu{\rm d}\Omega
\label{rate}
\end{equation}
The effective areas include all the details of the veto and absorption by the Earth
except for the atmospheric neutrino self-veto, which does not apply to astrophysical
neutrinos.  For example, for muon neutrinos below $60$~TeV, the acceptance is higher
for neutrinos from the Northern hemisphere because the atmospheric muon veto must be
more severe for events from above the detector (Southern hemisphere).  On the other
hand, at high energy the acceptance is smaller from below because of absorption of
neutrinos by the Earth.  

Overall, the acceptance of the HESE event selection is 
smallest for $\nu_\mu$ because of the starting track criterion coupled with the
high threshold for deposited energy in the detector and the muon veto.  At high
energy, much of the energy of a $\nu_\mu$-induced muon that starts in the
detector is deposited after the muon leaves the detector.  For
a flavor ratio of $1:1:1$ at Earth and for an astrophysical spectrum 
\begin{equation}
E_\nu^2\phi_\nu\,=\,1.5\times 10^{-8}\left(\frac{E_\nu}{100\,{\rm TeV}}\right)^{-0.3}\,\frac{{\rm GeV}}{{\rm cm}^2{\rm sr\,s}} \,{\rm per\,\,flavor},
\label{astrospectrum}
\end{equation}
the expected numbers of events in 988 days are $\nu_\mu:\nu_\tau:\nu_e\,\approx\,5:7:11$.
For comparison, the total number of events, including backgrounds in the 3-year 
analysis~\cite{Aartsen:2014gkd} is 37.  The astrophysical spectrum in Eq.~\ref{astrospectrum}
is a fit with a differential spectral index of $-2.3$ from Ref.~\cite{Aartsen:2014gkd}.

It is instructive to investigate the distributions of neutrino energies that
give rise to these astrophysical signals in comparison with the corresponding
distributions for the atmospheric backgrounds.  This comparison is shown in
Fig.~\ref{fig:nu-response}.  The plots show the number of events per logarithmic
interval of energy on a linear scale, so the area under each segment of
a curve correctly represents the fraction of events from that range of energy.
Another important point is that the plots show true neutrino energy
whereas the events are characterized by energy deposited in the detector.
Most of the atmospheric events have true neutrino energy less than $100$~TeV,
while most of the astrophysical signal is above that energy.
In calculating the rates of downward atmospheric neutrinos, the neutrino
fluxes at production have been reduced by the passing fraction as a function
of energy and zenith angle as described in Ref.~\cite{Gaisser:2014bja}.  A
consequence is that these backgrounds are expected to be smaller from the South 
than from the North, unlike the case for astrophysical neutrinos, where the
opposite is true.

The atmospheric neutrino numbers include the contribution of prompt
neutrinos calculated with the model of Enberg et al.~\cite{Enberg:2008te}
assuming a primary spectrum with the knee as in Ref.~\cite{Gaisser:2012zz}.
The conventional rates are calculated with the neutrino fluxes of the
energy-independent model discussed in Section~\ref{sec4} (the higher of
the two fluxes in the right panel of Fig.~\ref{fig:Zfactors}).  Most 
(7 of 8) of the expected atmospheric $\nu_\mu$ events are conventional (from 
decay of charged kaons and pions).  For atmospheric electron neutrinos
slightly more than half of an expected 3 events are prompt.  The expected
numbers of background neutrino events would be reduced proportionately
if the lower extrapolated flux in the right panel of Fig.~\ref{fig:Zfactors}
were used, but the ratios would be essentially unchanged.

The two panels in Fig.~\ref{fig:nu-response} show
$\nu_\mu$-induced events on the left and the 
sum of $\nu_\tau$ and $\nu_e$ on the right.
Most of the $\nu_\mu$ will be classified
as tracks, but some (including all neutral current interactions)
will be classified as cascades.  On the other hand all $\nu_e$
and most $\nu_\tau$ will be classified as cascades.  At present
the observations are consistent with a $1:1:1$ flavor ratio of
the astrophysical neutrinos at Earth.  The atmospheric neutrino
backgrounds should be mostly $\nu_\mu$-induced.  In principle,
given a significantly larger event sample and a good knowledge of 
 normalization of the background of atmospheric neutrinos at high energy, 
 it would be possible 
 to subtract the background and measure the astrophysical
flavor ratio.  At present, however,
the events are too few and the uncertainties in the atmospheric
background too large to do so.

\section{Summary and Outlook}
\label{sec7}

It is clearly important to acquire the best possible understanding
of the atmospheric neutrino fluxes at high energy.  The two main sources
of uncertainty are the primary spectrum and the limited knowledge of
hadronic interactions.  The biggest uncertainty at high energy
is the level of charm production.  In a paper at this conference~\cite{Felix},
work on a post-LHC model of the event generator SIBYLL was presented.
The new version includes production of charm and will therefore
give further insight into the level of prompt neutrinos.  Charm is
introduced in SIBYLL with a non-perturbative component, tuned to results
of fixed target experiments, for example Refs.~\cite{Alves:1996rz,Aitala:1996hf}
and a perturbative QCD component
tuned to LHC results, for example from ALICE~\cite{ALICE:2011aa} 
and LHCb~\cite{Aaij:2013mga}.

In Section~\ref{sec4} of this paper, we describe a method that can
be used to explore in a systematic way the implications of 
uncertainties in hadronic interactions and primary spectrum
for the fluxes of atmospheric neutrinos in the TeV-PeV energy range.

\section*{Acknowledgments}  I thank Dr. M. Honda for providing details of
the spectrum weighted moments for the hadronic interaction model used
to calculate the atmospheric neutrino fluxes in Ref.~\cite{Honda:2006qj}.
This research is supported in part by
grants from the U.S. National Science Foundation, NSF-PHY-1205809
and from the U.S. Department of Energy, 12ER41808.

\section*{Correction} 
The original version of this paper contains an important typographical error
in Eq.~\ref{an4}.  (The pre-factor on the r.h.s. was incorrectly written as $p-1$
rather than $p+1$.)  The corresponding code used to produce the energy-dependent
Z-factors was correct, so there are no changes in the results.  In particular,
the Z-factors and neutrino spectra in Fig.~\ref{fig:Zfactors} are correct and unchanged. 
I am grateful to Felipe Campos Penha for finding this error.

\end{document}